\documentclass{iopart}
\usepackage{iopams}

\newcommand{\Id}{{\mathrm{Id}}}

\newcommand{\perm}{{\mathrm{per}}}
\newcommand{\ee}{{\mathrm{e}}}
\newcommand{\dd}{{\mathrm{d}}}
\newcommand{\signe}{{\mathrm{sign}}}

\newcommand{\kbf}{{\mathbf{k}}}
\newcommand{\qbf}{{\mathbf{q}}}

\newcommand{\Fcal}{{\cal F}}
\newcommand{\Scal}{{\cal S}}
\newcommand{\Dcal}{{\cal D}}
\newcommand{\Acal}{{\cal A}}

\newcommand{\Jcal}{{\cal J}}

\renewcommand{\i}[1]{{}_{\scriptscriptstyle(#1)}}
\newcommand{\pvee}{{\scriptstyle\vee}}

\newcommand{\cou}{\varepsilon}

\newcommand{\antip}{\mathrm{s}}

\newcommand{\oprod}{\bullet}
 
\begin{document}

\title[Quantum groups and QFT. The free scalar field]
  { Quantum groups and quantum field theory: I. The free scalar field }
 
\author{Christian Brouder\dag\footnote[3]{(christian.brouder@lmcp.jussieu.fr)}
        and Robert Oeckl\ddag}
 
\address{\dag\ Laboratoire de Min\'eralogie-Cristallographie, CNRS UMR7590,
 Universit\'es Paris 6 et 7, IPGP, 4 place Jussieu,
  75252 Paris Cedex 05, France}
 
\address{\ddag\ Centre de Physique Th\'eorique,
   CNRS Luminy, F-13288 Marseille Cedex 9, France}
 
\begin{abstract}
The quantum field algebra of real scalar fields is shown 
to be an example of infinite dimensional quantum group.
The underlying Hopf algebra is the symmetric algebra $S(V)$
and the product is Wick's normal product.
Two coquasitriangular structures can be built from the
two-point function and the Feynman propagator 
of scalar fields to reproduce the operator product
and the time-ordered product as twist deformations of the
normal product. A correspondence is established between
the quantum group and the quantum field concepts.
On the mathematical side the
underlying structures come out of Hopf algebra cohomology.
\end{abstract}
 
\today
 


\maketitle

\section{Introduction}
This paper brings together quantum groups and quantum field theory.
More precisely, it shows that quantum field operator algebras are
infinite dimensional quantum groups.
A quantum group is a relatively recent mathematical structure that has
attracted considerable interest in mathematics and physics. 
It is a Hopf algebra equipped with a coquasitriangular
structure\footnote{More frequently, the quantum groups are defined
with a quasitriangular structure $R$ but the two notions
are dual to one another, see \cite{Majid}.}. 
Besides the quantum group structure for the field algebra,
many other concepts of quantum field theory can be translated
into the quantum group language. For instance, the normal product 
is the Hopf algebra product, the expectation value over the vacuum is
the counit, the operator product and time-ordered product are
twisted products.
 
This paper presents the basic concepts of quantum groups and their
applications to the free real scalar field. Fermions and 
interacting fields will be presented in forthcoming publications. 
We try to
familiarize the reader with the notions of Hopf algebra, Laplace
pairing, coquasitriangular structure and twisted product.
Results are derived in full detail.
 
In quantum field theory, the difference between a product of
two field operators and their normal product is
a scalar. 
Therefore, any product of field operators can be written as a linear 
combination of normal products of operators. 
We denote by $S(V)$ the vector space spanned by
all normal products of operators.
Similarly, the difference between the time-ordered product
of two operators and their normal product
is a scalar function. Thus, all time-ordered
products of operators belong also to $S(V)$.
In other words, quantum field theory can be entirely
discussed in the space $S(V)$ of normal products.
Working in $S(V)$ has several advantages, already
noted by Houriet and Kind \cite{Houriet} and later
stressed by Wick \cite{Wick} and by Gupta
\cite{Gupta}: $S(V)$ is equipped with a commutative
product (the normal product) and the
expectation value of an element $u$ of $S(V)$ over the vacuum
is zero if $u$ has no scalar component.
 
This motivates us to use the Hopf algebra structure of $S(V)$. The
product of $S(V)$ is the normal product while the counit describes
the vacuum expectation value. Remarkably, both the operator product
and the time-ordered product arise now as twist deformations of the
normal ordered product.
More precisely, we first make $S(V)$ into a quantum
group by adding a Laplace pairing (or coquasitriangular 
structure). The Laplace pairing is determined by
appropriate Green functions or propagators. This Laplace pairing
then gives rise to a new product on $S(V)$, called
\emph{twisted product} (a special case of Sweedler's crossed product
\cite{Sweedler}), which turns out to reproduce the operator or the 
time-ordered product, depending on the chosen Green function.
In this sense, quantum groups enable us to define second quantization
without commutators.
For the time-ordered product we also obtain the time
ordering prescription $T$ linking the normal and time-ordered product
as an algebra automorphism $T:S(V)\to S(V)$.
 
Note that our approach here is closely related to the one of braided
quantum field theory \cite{Oeckl} which employs a path integral
picture. The
starting point there is the free tensor algebra over the
classical fields to be compared with $S(V)$ in our context. (There,
the analog of our coquasitriangular structure is the map $\gamma$.)
Furthermore, the connection between a twisted product and Wick's
theorem was already proposed by Fauser in a
fermionic (i.e.\ antisymmetric) setting \cite{Fauser}.
In general, the quantum group framework
enables us to replace combinatorial proofs of 
quantum field theory by algebraic manipulations.
Several examples of this were given in 
reference \cite{BrouderOeckl}.
 
This paper is broadly divided into two parts. In the first part
(Sections~\ref{sec:symqg} and \ref{sec:twist}) we introduce the required
mathematical
structures while the second applies them to quantum field theory
(Section~\ref{sec:qft}). We start in Section~\ref{sec:symqg} with the
definition of the Hopf algebra $S(V)$ and a Laplace pairing.
In Section~\ref{sec:twist} we proceed to introduce the
twisted product on $S(V)$, based on the Laplace pairing.
Furthermore, in the case where the twisted product is commutative, we
construct the algebra automorphism $T:S(V)\to S(V)$ which relates the
twisted and untwisted products. We also show that this $T$ can be
expressed as an exponential. In Section~\ref{sec:qft} we turn to
quantum field theory. First, we conveniently define the vector space
$V$ of smoothed field operators on a Fock space. $S(V)$ is then
identified as the Hopf algebra of polynomials in the operators with
normal product. Using appropriate Green functions we define
Laplace pairings which give rise to the
operator and time-ordered products. The proof that we indeed obtain
these products as twisted products is performed by invoking Wick's
theorem. For the time-ordered product the $T$-map gives a compact
expression for the vacuum expectation value of the free field theory.
Finally, the conclusion describes some prospects of the present work.
The appendix provides a lightning introduction to Hopf algebras.

\section{The symmetric quantum group}
\label{sec:symqg}
 
A quantum group is a Hopf algebra with a (co)quasitriangular structure
and we describe now the Hopf algebra that we need in the
scalar quantum field theory: the symmetric Hopf algebra $S(V)$.
Readers who are not familiar with Hopf
algebras should first read the appendix.
 
\subsection{The symmetric Hopf algebra}
\label{sec:symHopf}
 
Later, the symmetric product
will be used to describe the normal product of bosonic creation
and annihilation operators (see Section~\ref{normalprodsect})
but we define now the symmetric Hopf algebra for
a general vector space $V$ over $\mathbb{C}$.
The symmetric algebra $S(V)$ is the direct sum
\begin{eqnarray*}
S(V) &=& \bigoplus_{n=0}^\infty S^n(V),
\end{eqnarray*}
where $S^0(V)=\mathbb{C}$ and $S^1(V)=V$.
For any vector space $V$, $S(V)$ is 
the free unital commutative and associative algebra over $V$
\cite{Eisenbud}. If $V$ is finite dimensional with basis $e_i$,
$S^n(V)$ is spanned by elements of the
form $e_{i_1}\pvee\cdots\pvee e_{i_n}$,
with $i_1 \leq i_2 \leq\dots\leq i_n$.
The symbol $\pvee$ denotes the associative and commutative product
$\pvee : S^m(V) \otimes S^n(V) \longrightarrow S^{m+n}(V)$.
For finite dimensional $V$, the product is defined on monomials by
$(e_{i_1}\pvee\cdots\pvee e_{i_m})\pvee (e_{i_{m+1}}\pvee\cdots\pvee 
e_{i_{m+n}}) = e_{i_{\sigma(1)}}\pvee\cdots\pvee e_{i_{\sigma(m+n)}}$,
where $\sigma$ is the permutation on $m+n$ elements such that
$i_{\sigma(1)}\leq\cdots\leq i_{\sigma(m+n)}$
and then extended by linearity and associativity
to all elements of $S(V)$. The unit of $S(V)$ is the scalar unit
$1\in S^0(V)$ (i.e.\ for any $u\in S(V)$: $1\pvee u=u\pvee 1=u$).
The algebra $S(V)$ is graded, the degree of an element
$u$ of $S^n(V)$ is $|u|=n$. 
In the finite dimensional case, $S(V)$ can be seen as the algebra 
of polynomials
in the variables $\{e_i\}$, the elements of
$S^n(V)$ being homogeneous polynomials of degree $n$.
 
In this paper, $u$, $v$, $w$ designate elements of $S(V)$,
$a$, $a_i$, $b$, $c$ elements of $V$.
 
The coproduct of $S(V)$ is defined as follows
\begin{eqnarray}
\Delta 1 &=& 1\otimes 1,\nonumber\\
\Delta a &=& a\otimes 1 + 1\otimes a\quad{\mathrm{for}}\quad a\in V,
\label{Deltaa}\\
\Delta (u\pvee v)&=& \sum (u\i1\pvee v\i1) \otimes (u\i2\pvee
v\i2),\label{Deltau}
\end{eqnarray}
where Sweedler's notation was used for the coproduct of $u$ and $v$:
$\Delta u=\sum u\i1\otimes u\i2$ and
$\Delta v=\sum v\i1\otimes v\i2$.
This coproduct is coassociative and cocommutative.
As an example, we calculate the coproduct $\Delta (a\pvee b)$,
with $a$ and $b$ in $V$.
From equation (\ref{Deltau})
\begin{eqnarray*}
\Delta (a\pvee b)&=& \sum (a\i1\pvee b\i1) \otimes (a\i2\pvee b\i2),
\end{eqnarray*}
From the definition (\ref{Deltaa}) of the coproduct acting on elements
of $V$ we know that
$\Delta a = a\otimes 1 + 1\otimes a$
and
$\Delta b = b\otimes 1 + 1\otimes b$.
Thus, 
\begin{eqnarray*}
\Delta (a\pvee b)&=& \sum (a\i1\pvee b\i1) \otimes (a\i2\pvee b\i2)\\
&=& (a\pvee b)\otimes (1\pvee 1) +
    (a \pvee 1)\otimes (1\pvee b) +
    (1 \pvee b) \otimes (a \pvee 1) 
\\&& + (1 \pvee 1) \otimes (a \pvee b)
\\&=&
(a\pvee b)\otimes 1 + a \otimes  b + b \otimes a + 1 \otimes (a \pvee b).
\end{eqnarray*}
At the next degree, we have
\begin{eqnarray*}
\Delta (a\pvee b\pvee c) &=&
 1\otimes a\pvee b\pvee c
 +a \otimes b\pvee c
 +b \otimes a\pvee c 
 +c \otimes a\pvee b
\\&&
 +a\pvee b \otimes c
 +a\pvee c \otimes b
 +b\pvee c \otimes a
+a\pvee b\pvee c\otimes 1.
\end{eqnarray*}
 
Generally, if
$u=a_1\pvee a_2\pvee \dots \pvee a_n$,
the coproduct of $u$ 
can be written explicitly as (\cite{Loday98} p.450)
\begin{eqnarray}
\Delta u &=& u\otimes 1 + 1 \otimes u
\nonumber\\&&+
\sum_{p=1}^{n-1} \sum_{\sigma}
a_{\sigma(1)}\pvee\dots\pvee a_{\sigma(p)}
\otimes a_{\sigma(p+1)}\pvee\dots\pvee a_{\sigma(n)},
\label{shuffle}
\end{eqnarray}
where $\sigma$ runs over the $(p,n-p)$-shuffles.
A $(p,n-p)$-shuffle is a permutation $\sigma$ of
$(1,\dots,n)$ such that $\sigma(1)<\sigma(2)<\dots <\sigma(p)$
and $\sigma(p+1)<\dots <\sigma(n)$.
 
The counit is defined by $\varepsilon(1)=1$ and
$\varepsilon(u)=0$ if $u\in S^n(V)$ and $n>0$.
The antipode is defined by $\antip(u)=(-1)^n u$ if $u\in S^n(V)$.
In particular, $\antip(1)=1$. Since the symmetric product
is commutative, the antipode is an algebra homomorphism:
$\antip(u\pvee v)=\antip(u) \pvee \antip(v)$. 
 
The fact that this coproduct is coassociative
and that $\Delta$, $\varepsilon$ and $S$ give $S(V)$ 
the structure of a cocommutative Hopf algebra, are classical results
(see \cite{Eisenbud,Loday98,Kassel} and references therein).

\subsection{The Laplace pairing}
 
We define a \emph{Laplace pairing} on $S(V)$ as a
bilinear form $V\times V\rightarrow \mathbb{C}$,
that we denote by $(a|b)$, which is extended to $S(V)$ by 
$(1|1)=1$ and the following recursions
\begin{eqnarray}
(u\pvee v|w) &=& \sum (u|w\i1) (v|w\i2), \label{laplaceid1}\\
(u | v\pvee w) &=& \sum (u\i1|v) (u\i2|w).\label{laplaceid2}
\end{eqnarray}
It is important to stress that we do not assume any special symmetry
for the bilinear form (i.e.\ it is a priori neither symmetric nor
antisymmetric).
 
These
recursions have the following unique solution \cite{Grosshans}:
If $u=a_1\pvee a_2\pvee \dots \pvee a_k$ and
$v=b_1\pvee b_2\pvee \dots \pvee b_n$, then
$(u|v)=0$ if $n\not=k$ and
$(u|v)=\perm(a_i|b_j)$ if $n=k$.
We recall that the {\emph{permanent}} of the matrix  $(a_i|b_j)$ is
\begin{eqnarray}
\perm(a_i|b_j) &=& \sum_{\sigma} (a_1|b_{\sigma(1)}) \cdots
(a_k|b_{\sigma(k)}),
\label{permanent}
\end{eqnarray}
where the sum is over all permutations $\sigma$ of $\{1,\dots,n\}$.
The permanent is a kind of determinant where all signs are positive.
For instance
\begin{eqnarray*}
(a\pvee b| c\pvee d) &=& (a|c)(b|d) + (a|d)(b|c).
\end{eqnarray*}
 
An important concept in the theory of quantum groups is that of a
\emph{coquasitriangular structure} \cite{Majid}.
This is a bilinear form
on a Hopf algebra satisfying certain conditions.
Indeed one may define a
quantum group precisely as a Hopf algebra with a coquasitriangular
structure.
The conditions coincide in the case of $S(V)$ with that of a
Laplace pairing, i.e.\ the two concepts are identical for $S(V)$. 
The terminology of these concepts is not settled yet in
the Hopf algebra literature. A coquasitriangular structure
is also called an $R$-form \cite{Jacobs},
a braiding \cite{Hodges}
or a universal $r$-form \cite{Schmudgen},
a Laplace pairing\footnote{For a general Hopf algebra,
equation (\ref{laplaceid2}) becomes
$(u | v\pvee w) = \sum (u\i2|v) (u\i1|w)$
in the definition of a Laplace pairing.
This is equivalent to (\ref{laplaceid2}) in our case
of a cocommutative Hopf algebra.}
is also called
a duality pairing \cite{Majid}, 
a skew pairing \cite{Hodges}, a dual pairing \cite{Majid2002},
a bicharacter \cite{Bahturin} 
or a Hopf pairing \cite{BrownGoodearl}.
We use the term ``Laplace pairing'' because it was introduced 
very early by Rota and coll.\ in \cite{DRS} and it reflects
the fact that equations (\ref{laplaceid1}) and (\ref{laplaceid2}) 
are an elegant way of writing the Laplace identities
for permanents \cite{Minc}. In the case of determinants, 
the Laplace identities express the determinant in terms
of a sum of products of minors 
(see reference \cite{Muir} p.93). 
 
As an exercise, we derive the following identity:
\begin{eqnarray}
(u|1)= (1|u)=\varepsilon(u).
\label{unu}
\end{eqnarray}
Because of the definition of the counit and the antipode we have
\begin{eqnarray*}
u &=& \sum \varepsilon(u\i1) u\i2
= \sum \antip(u\i1\i1) \pvee u\i1\i2 \pvee u\i2.
\end{eqnarray*}
From the fact that $\Delta 1 = 1\otimes 1$, we deduce
from equation (\ref{laplaceid1})
\begin{eqnarray*}
(u|1) &=& \sum (\antip(u\i1\i1)|1) (u\i1\i2|1) (u\i2|1)
\\ &=& \sum (\antip(u\i1)|1) (u\i2\i1|1) (u\i2\i2|1)
= \sum (\antip(u\i1)|1) (u\i2|1\pvee 1)
\\  
&=& \sum (\antip(u\i1)u\i2|1\pvee 1) = \varepsilon(u) (1|1) = \varepsilon(u).
\end{eqnarray*}


\section{The twisted product}
\label{sec:twist}
 
This section introduces the \emph{twisted product}. 
Its importance stems form the fact that it will allow us to
treat in one single stroke the operator product and
the time-ordered product, depending on the definition
of the bilinear form $(a|b)$. We define it as
\begin{eqnarray}
u\circ v &=& \sum u\i1\pvee v\i1 \, (u\i2|v\i2).
\label{deftwisted}
\end{eqnarray}
This product (for general commutative Hopf algebras) was
introduced by Sweedler as a \emph{crossed
product} in Hopf algebra cohomology theory \cite{Sweedler}. Indeed, it
turns out that a Laplace pairing in our case is not only a
coquasitriangular structure but also a 2-cocycle. Drinfeld showed that
2-cocycles give rise to \emph{twist deformations} of Hopf algebras and
their representation categories \cite{Drinfeld}. The twisted product
can be understood as precisely such a twist deformation\footnote{To
be precise, the twisted product is a twist deformation as a comodule
algebra, not as a Hopf algebra, see \cite{OecklPhD}.}.
As we aim to be introductory here we shall not dwell on these facts
although they are the conceptual origin of much of this section
including the $T$-map. More details will be given in 
\cite{BrouderQG}.
 
 
By cocommutativity of the coproduct on $S(V)$, the definition
(\ref{deftwisted}) is equivalent to
\begin{eqnarray*}
u\circ v &=& \sum  u\i1\pvee v\i2 \, (u\i2|v\i1)
= \sum  u\i2\pvee v\i1 \, (u\i1|v\i2)
\\&=&
\sum  (u\i1|v\i1)\, u\i2\pvee v\i2.
\end{eqnarray*}
We start with the simplest example. If $a$ and $b$ are in $V$,
\begin{eqnarray*}
a\circ b &=& (1|1) a\pvee b + (a|1) 1\pvee b + (1|b) a\pvee 1
  +(a|b) 1\pvee 1
\\&=& a\pvee b + (a|b) 1.
\end{eqnarray*}
For notational convenience, we do not write the algebra
unit explicitly in the rest of the paper.
Thus $a\circ b = a\pvee b + (a|b)$. 
A few more examples might be useful.
\begin{eqnarray*}
(a\pvee b)\circ c &=& a\pvee b \pvee c +(a|c)b + (b|c) a, \\
a\circ ( b \pvee c) &=& a\pvee b \pvee c +(a|c)b + (a|b) c, \\
a\circ b \circ c &=& a\pvee b \pvee c + (a|b)c +(a|c)b + (b|c) a, \\
a\circ b \circ c \circ d &=& a\pvee b \pvee c\pvee d + 
(a|b)c\pvee d +(a|c)b\pvee d 
\\&&\hspace*{-0mm}
+ (b|c) a\pvee d
+(a|d)b\pvee c +(b|d)a\pvee c 
+ (c|d) a\pvee b 
\\&&\hspace*{-0mm}
+(a|b)(c|d) +(a|c)(b|d) + (b|c) (a|d), \\
(a\pvee b) \circ (c \pvee d) &=& 
a\pvee b \pvee c\pvee d 
+(a|c)b\pvee d + (b|c) a\pvee d
\\&&\hspace*{-0mm}
+(a|d)b\pvee c +(b|d)a\pvee c 
+(a|c)(b|d) + (b|c) (a|d).
\end{eqnarray*}
We shall need some simple properties of the twisted product
\begin{eqnarray*}
u\circ 1 &=& 1\circ u = u, \\
u\circ (v+w) &=& u\circ v+u\circ w, \\
u\circ (\lambda v) &=& (\lambda u)\circ v= \lambda (u\circ v).
\end{eqnarray*}
 
As an exercise, we prove that
\begin{eqnarray}
(u|v) &=& \varepsilon(u\circ v). \label{circvarepsilon}
\end{eqnarray}
We apply the counit $\varepsilon$ to definition (\ref{deftwisted})
of the twisted product:
\begin{eqnarray*}
\varepsilon(u\circ v) &=& \sum \varepsilon(u\i1\pvee v\i1) (u\i2|v\i2).
\end{eqnarray*}
The counit is an algebra homomorphism, i.e.
$\varepsilon(u\i1\pvee v\i1)=\varepsilon(u\i1)\varepsilon(v\i1)$.
Thus, by linearity of the Laplace pairing,
\begin{eqnarray*}
\varepsilon(u\circ v) &=& \sum (\varepsilon(u\i1)u\i2|\varepsilon(v\i1)v\i2).
\end{eqnarray*}
The counit property is precisely the identity
$\sum \varepsilon(u\i1)u\i2 = u$. This proves equation (\ref{circvarepsilon}).

\subsection{Proof of associativity}
 
The twisted product is associative \cite{Sweedler}: $u\circ (v\circ w)
= (u\circ v)\circ w$. Indeed, this follows from the fact that it is a
twist
deformation. However, as the associativity is an important result, we
provide here for the convenience of the reader an alternative
self-contained proof.
 
We first need a useful lemma:
\begin{eqnarray}
\Delta (u\circ v) &=& \sum (u\i1\pvee v\i1) \otimes (u\i2\circ v\i2)
     \label{Deltacirc} \\
&=& \sum (u\i1\circ v\i1)\otimes (u\i2\pvee v\i2).
     \label{Deltacirc2}
\end{eqnarray}
The proof is easy, we start from the definition of the twisted product
(\ref{deftwisted})  to write
\begin{eqnarray*}
\Delta (u\circ v) &=& \sum \Delta(u\i1\pvee v\i1) (u\i2| v\i2).
\end{eqnarray*}
Now we use the definition (\ref{Deltau}) of the coproduct of $u\i1\pvee v\i1$:
\begin{eqnarray*}
\Delta (u\circ v) &=& \sum (u\i1\i1\pvee v\i1\i1)\otimes
(u\i1\i2\pvee v\i1\i2) (u\i2| v\i2).
\end{eqnarray*}
The coassociativity of the coproduct of $u$ means that any triple 
$(u\i1\i1,u\i1\i2,u\i2)$ can be replaced by
$(u\i1,u\i2\i1,u\i2\i2)$. Therefore,
\begin{eqnarray*}
\Delta (u\circ v) &=& \sum (u\i1\pvee v\i1\i1)\otimes (u\i2\i1\pvee
v\i1\i2) (u\i2\i2| v\i2). 
\end{eqnarray*}
We use now the coassociativity of the coproduct of $v$
\begin{eqnarray*}
\Delta (u\circ v) &=& \sum (u\i1\pvee v\i1)\otimes (u\i2\i1\pvee
v\i2\i1) (u\i2\i2| v\i2\i2)
\end{eqnarray*}
and the definition of the twisted product brings
\begin{eqnarray*}
\Delta (u\circ v) &=& \sum (u\i1\pvee v\i1)\otimes (u\i2\circ v\i2).
\end{eqnarray*}
To obtain the other identity of the lemma, we start from 
$u\circ v=\sum (u\i1| v\i1) u\i2\pvee v\i2$ or we use the
cocommutativity of the coproduct.
 
The second lemma is
\begin{eqnarray}
(u|v\circ w) &=& (u\circ v|w). \label{lemma2}
\end{eqnarray}
The proof is straightforward
\begin{eqnarray*}
(u|v\circ w) &=& \sum (u|v\i1\pvee w\i1) (v\i2| w\i2)
\\&=&
\sum (u\i1|v\i1)(u\i2| w\i1) (v\i2| w\i2)
\\&=&
\sum (u\i1|v\i1)(u\i2\pvee v\i2| w)
=(u\circ v|w).
\end{eqnarray*}
The first line is the definition of the twisted product
(\ref{deftwisted}), the second line is
the expansion of the Laplace pairing (\ref{laplaceid2}), the third
one is the Laplace identity (\ref{laplaceid1}) and the 
last equality is again equation (\ref{deftwisted}).
 
The associativity of the twisted product follows immediately from
these two lemmas. From definition (\ref{deftwisted}) and the
first lemma we obtain
\begin{eqnarray*}
u\circ(v\circ w) &=&
\sum u\i1\pvee (v\circ w)\i1\,(u\i2| (v\circ w)\i2)
\\&=&
 \sum u\i1\pvee (v\i1\pvee w\i1)\,(u\i2|v\i2\circ w\i2).
\end{eqnarray*}
From the associativity of the symmetric product and the second lemma
we obtain
\begin{eqnarray*}
u\circ(v\circ w) &=&
\sum u\i1\pvee v\i1\pvee w\i1\,(u\i2\circ v\i2| w\i2).
\end{eqnarray*}
Now the first lemma enables us to rewrite this as
\begin{eqnarray*}
u\circ(v\circ w) &=&
\sum (u\circ v)\i1\pvee w\i1\,((u\circ v)\i2| w\i2)
= (u\circ v)\circ w.
\end{eqnarray*}
 

\subsection{Inversion of twisted product}
 
The fact that the twisted product is defined through a twist deformation
means that it can be inverted. That is, the symmetric product can be
written in terms of the twisted product. Indeed, the formula is very
simple and uses the antipode (see e.g.\ \cite{Majid,RotaStein94}:
\begin{eqnarray*}
u\pvee v &=& \sum (\antip(u\i1)|v\i1) u\i2\circ v\i2
= \sum (u\i1|\antip(v\i1)) u\i2\circ v\i2.
\label{symmetric-twisted}
\end{eqnarray*}
The proof is simple:
\begin{eqnarray*}
\sum (\antip(u\i1)|v\i1) u\i2\circ v\i2 &=&
\sum (\antip(u\i1)|v\i1) (u\i2\i1|v\i2\i1) u\i2\i2\pvee v\i2\i2
\\&=&
\sum (\antip(u\i1\i1)|v\i1\i1) (u\i1\i2|v\i1\i2) u\i2\pvee v\i2
\\&=&
\sum (\antip(u\i1\i1)\pvee u\i1\i2|v\i1) u\i2\pvee v\i2
\\&=&
\sum \varepsilon(u\i1)(1 |v\i1) u\i2\pvee v\i2
\\&=&
\sum \varepsilon(u\i1)\varepsilon(v\i1) u\i2\pvee v\i2 = u\pvee v.
\end{eqnarray*}
The last line was obtained because of (\ref{unu}).
It is also possible to recover the Laplace
pairing from the twisted product, as already shown in
(\ref{circvarepsilon}).
Reference \cite{RotaStein94} also provides a sort of distributivity
formula for twisted and symmetric products:
\begin{eqnarray}
u \circ (v\pvee w) &=& \sum (u\i1\circ v)
\pvee\antip(u\i2\i1) \pvee (u\i2\i2\circ w)
\label{distributivity}
\end{eqnarray}
For this result we need the following lemma
\begin{eqnarray*}
\sum s(u\i1)\pvee (u\i2\circ v) &=& \sum v\i1 (u | v\i2)
\end{eqnarray*}
which is proved easily
\begin{eqnarray*}
\sum s(u\i1)\pvee (u\i2\circ v) &=& \sum 
s(u\i1) \pvee u\i2\i1\pvee v\i1(u\i2\i2 | v\i2)   
\\&=&
\sum s(u\i1\i1) \pvee u\i1\i2\pvee v\i1(u\i2 | v\i2) 
\\&=&
\sum \varepsilon(u\i1) v\i1(u\i2 | v\i2)
=
\sum v\i1 (u | v\i2).
\end{eqnarray*}
Now, we can prove (\ref{distributivity}).
Using the last lemma, we obtain
\begin{eqnarray*}
\sum (u\i1\circ v) \pvee\antip(u\i2\i1) \pvee (u\i2\i2\circ w)
 &=& 
\sum (u\i1\circ v) \pvee w\i1 (u\i2| w\i2).
\end{eqnarray*}
On the other hand, we have
\begin{eqnarray*}
\sum u \circ (v\pvee w)
&=&\sum u\i1\pvee v\i1\pvee w\i1 (u\i2|v\i2\pvee w\i2)
\\
 &=& \sum u\i1\pvee v\i1\pvee w\i1 (u\i2\i1|v\i2)(u\i2\i2|w\i2)
\\
 &=& \sum u\i1\i1\pvee v\i1\pvee w\i1 (u\i1\i2|v\i2)(u\i2|w\i2)
\\
&=&
\sum (u\i1\circ v) \pvee w\i1 (u\i2| w\i2).
\end{eqnarray*}
This proves the lemma.
 
In the proofs
of the lemmas of the present section
and of the coassociativity of the twisted product,
we never used the commutativity of the symmetric product. 
Therefore, all our results are still valid if we work in the tensor
algebra $T(V)$ instead of the symmetric algebra $S(V)$.
 
\subsection{Commutative twisted product}
In quantum field theory, the time-ordered product is not
considered as a product but as an operation 
$T(\varphi_1 \cdots \varphi_n)$ that orders the
operator product of $\varphi_1$, \dots, $\varphi_n$
according to decreasing times
(see, e.g.\ \cite{Itzykson} p.177).
Since the arguments of $T$ commute, we can consider
$T$ as a map from $S(V)$ to $S(V)$.
Moreover, we shall show that the time-ordered product can
indeed be considered as a commutative twisted product. 
Thus, from now on in this section, the twisted product 
will be considered commutative.
We first show that this is equivalent to the condition $(a|b)=(b|a)$
for all $a,b\in V$.
 
If the twisted product is commutative, then
in particular $a\circ b=b\circ a$, so that 
$a\pvee b+ (a|b)=b\pvee a+(b|a)$ and $(a|b)=(b|a)$.
Conversely, if $(a|b)=(b|a)$ for all $a,b\in V$, 
then $u\circ v=v\circ u$ for all $u,v \in S(V)$.
To prove the last statement, we first show that, if $(a|b)=(b|a)$
for all $a,b\in V$, then $(u|v)=(v|u)$ for all $u,v \in S(V)$.
This follows immediately from the definition (\ref{permanent}) of
the permanent.
Because of the commutativity of the symmetric product $\pvee$
we obtain
\begin{eqnarray*}
u\circ v &=& \sum u\i1\pvee v\i1 \, (u\i2|v\i2)
=\sum v\i1\pvee u\i1 \, (v\i2|u\i2) = v \circ u.
\end{eqnarray*}
 
\subsection{The T-map}
We define recursively a linear map $T$ from $S(V)$ to $S(V)$ by
$T(1)=1$, $T(a)=a$ for $a$ in
$V$ and $T(u\pvee v)=T(u)\circ T(v)$.
More explicitly, $T(a_1\pvee\dots\pvee a_n) = a_1\circ\dots\circ a_n$.
The twisted product is associative and commutative,
therefore $T$ is well defined on $S(V)$.
This T-map is the usual time-ordering operator of quantum field theory.
 
The main property we shall need in the following is
\begin{eqnarray}
\Delta T(u) &=& \sum u\i1 \otimes T(u\i2)= \sum T(u\i1) \otimes u\i2.
\label{DeltaT}
\end{eqnarray}
We use a recursive argument. Equation (\ref{DeltaT}) is true
for $u=1$ or $u=a$ and $v=1$ or $v=b$.
Now
\begin{eqnarray*}
\Delta T(u\pvee v) &=& \Delta \big( T(u)\circ T(v)\big).
\end{eqnarray*}
By equation (\ref{Deltacirc}) we obtain
\begin{eqnarray*}
\Delta T(u\pvee v) &=& \sum \big(T(u)\i1\pvee T(v)\i1\big)
\otimes
\big(T(u)\i2\circ T(v)\i2\big).
\end{eqnarray*}
The recursion hypothesis yields
\begin{eqnarray*}
\Delta T(u\pvee v) &=& \sum \big(u\i1\pvee v\i1\big)
\otimes
\big(T(u\i2)\circ T(v\i2)\big)
\\&=&
\sum (u\pvee v)\i1
\otimes
T(u\i2\pvee v\i2).
\end{eqnarray*}
This gives us the expected result. The symmetric
identity in equation (\ref{DeltaT}) is obtained by
cocommutativity of the coproduct.

\subsection{Exponentiation}
 
Now, we shall derive a result obtained by Anderson \cite{Anderson}
and rediscovered several times \cite{Stumpf}:
the T-map can be written as the exponential of an operator $\Sigma$.
As an application of the Laplace Hopf algebra, we give a purely
algebraic proof of this. For simplicity, we consider a
finite dimensional vector space $V$.
 
First, we define a derivation $\delta_k$ attached to
a basis $\{e_k\}$ of the vector space $V$. This derivation is defined
as the linear operator on $S(V)$ satisfying the following properties
\begin{eqnarray}
\delta_k 1 &=& 0,\nonumber\\
\delta_k e_j &=& \delta_{kj},\label{deriv1}\\
\delta_k (u\pvee v) &=& (\delta_k u)\pvee v + u \pvee (\delta_k v).
\label{deriv2}
\end{eqnarray}
In the second equation, $\delta_{kj}$ is 1 for $j=k$ and
zero otherwise.
The Leibniz relation (\ref{deriv2}) gives us
$\delta_1 (e_1\pvee e_2)=e_2$ and
$\delta_2 (e_1\pvee e_2)=e_1$, for example.
From this definition, it can be shown recursively 
that the derivatives commute: $\delta_i\delta_j=\delta_j\delta_i$.
Note that the derivation does not act on the Laplace pairing:
$\delta_i (u|v) =0$.
 
Now we define the infinitesimal T-map as
\begin{eqnarray*}
\Sigma &=& \frac{1}{2} \sum_{ij} (e_i|e_j) \delta_i \delta_j,
\end{eqnarray*}
where the sum is over all elements of the basis of $V$.
 
We proceed in several steps to show that $T=\exp\Sigma$.
First, we show that 
\begin{eqnarray}
[\Sigma, a] &=& \sum_i (a|e_i) \delta_i. \label{Sigmaa}
\end{eqnarray}
To do that, we apply $\Sigma$ to an element $e_k\pvee u$
\begin{eqnarray*}
\Sigma (e_k \pvee u) &=& \frac{1}{2} \sum_{ij} (e_i|e_j) \delta_i
\delta_j (e_k \pvee u)
\\&=&
\frac{1}{2} \sum_{ij} (e_i|e_j) \delta_i (\delta_{jk} u + e_k \pvee \delta_j u)
\\&=&
\frac{1}{2} \sum_{i} (e_i|e_k) \delta_i u
+\frac{1}{2} \sum_{j} (e_k|e_j) \delta_j u
+\frac{1}{2} \sum_{ij} (e_i|e_j) e_k \pvee \delta_i \delta_j u
\\&=& \sum_{j} (e_k|e_j) \delta_j u +e_k \pvee \Sigma u.
\end{eqnarray*}
If we extend this by linearity to $V$ we obtain
\begin{eqnarray}
\Sigma(a\pvee u)=a\pvee \Sigma u+\sum (a|e_j) \delta_j u. \label{Sigmaakwu}
\end{eqnarray}
And more generally
\begin{eqnarray*}
\Sigma (u \pvee v) &=& (\Sigma u) \pvee v + u \pvee (\Sigma v)
+ \sum_{ij} (e_i|e_j) (\delta_i u )\pvee (\delta_j v).
\end{eqnarray*}
 
From the commutation of the derivations we obtain
$[\Sigma,\delta_k]=0$ and $[\Sigma,[\Sigma, a]]=0$.
Therefore, the classical formula (\cite{Itzykson} p.167) yields
\begin{eqnarray*}
\ee^\Sigma a \ee^{-\Sigma} &=& a + [\Sigma,a],
\end{eqnarray*}
so that
\begin{eqnarray*}
[\ee^\Sigma,a] &=& \sum_i (a|e_i) \delta_i \ee^\Sigma = 
\ee^\Sigma \sum_i (a|e_i) \delta_i.
\end{eqnarray*}
This can be written more precisely as
\begin{eqnarray}
\ee^\Sigma (a\pvee u) &=& a\pvee (\ee^{\Sigma}u) +  [\Sigma,a]
(\ee^{\Sigma}u). \label{eSigmaaveeu}
\end{eqnarray}
 
In the course of the proof of Wick's theorem, we shall derive
equation (\ref{ucircb}) which can be rewritten
\begin{eqnarray}
a\circ u &=& a\pvee u + [\Sigma,a]u. \label{acircu}
\end{eqnarray}
Now we have all we need to prove inductively that $T=\ee^\Sigma$.
We have $T(1)=\ee^\Sigma 1=1$ and
$T(a)=a=\ee^\Sigma a$ because $\Sigma a=0$.
Now assume that the property is true up to degree $k$, we
take an element $u$ of $S^k(V)$ and calculate 
\begin{eqnarray*}
T(a\pvee u) &=& a \circ T(u) = a\pvee T(u) + [\Sigma,a] T(u)
\\&=& a \pvee \ee^\Sigma u + [\Sigma,a] \ee^\Sigma u= \ee^\Sigma (a\pvee u),
\end{eqnarray*}
where we used equation (\ref{acircu}), then equation (\ref{eSigmaaveeu}).
Thus the property is true for $a\pvee u$ whose degree is $k+1$.
 
\subsection{The scalar t-map}
\label{sec:tmap}
 
It is possible to write the T-map as a sum of scalars multiplied by
elements of $S(V)$. In fact we can show that
\begin{eqnarray}
T(u) &=& \sum t(u\i1) u\i2, \label{Tt}
\end{eqnarray}
where the map $t$ is a linear map from $S(V)$ to $\mathbb{C}$,
called the t-map and defined recursively by
$t(1)=1$, $t(a)=0$ for $a\in V$ and
\begin{eqnarray}
t(u\pvee v) &=& \sum t(u\i1) t(v\i1) (u\i2 | v\i2).
\label{deft}
\end{eqnarray}
The proof is recursive. If the property is true
up to degree $k$, we take $w=u\pvee v$, where $u$ and $v$ have
degree $k$ or smaller and we calculate
\begin{eqnarray*}
T(w) &=& T(u)\circ T(v) = \sum t(u\i1) t(v\i1) u\i2 \circ v\i2
\\&=&
\sum t(u\i1) t(v\i1) (u\i2\i1|v\i2\i1) u\i2\i2 \pvee v\i2\i2
\\&=&
\sum t(u\i1\i1) t(v\i1\i1) (u\i1\i2|v\i1\i2) u\i2 \pvee v\i2
\\&=&
\sum t(u\i1\pvee v\i1) u\i2 \pvee v\i2 =\sum t(w\i1) w\i2,
\end{eqnarray*}
where the first line is the definition of $T(u\pvee v)$ and the 
recursion hypothesis, the second line is the definition of
the twisted product, the third line is the coassociativity of the
coproduct and the last line is the definition of $t$.
 
Notice that the t-map is well defined because
$t(u)=\varepsilon(T(u))$:
\begin{eqnarray*}
\varepsilon(T(u)) 
&=&
\sum t(u\i1) \varepsilon(u\i2)
=
t\big( \sum u\i1 \varepsilon(u\i2)\big) = t(u).
\end{eqnarray*}
 
The scalar $t$-map can be calculated from its definition
(\ref{deft}). The first degrees give
\begin{eqnarray*}
t(a \pvee b) &=& (a|b),\\
t(a \pvee b \pvee c \pvee d) &=& (a|b)(c|d) + (a|c)(b|d) +(a|d)(b|c).
\end{eqnarray*}
In general, $t(u)=0$ if $|u|$ is odd
and $t(u)$ is given by the following formula
if $u=a_1 \pvee \dots \pvee a_{2n}$ :
\begin{eqnarray}
t(u) &=& 
\sum_\sigma \prod_{k=1}^{n}(a_{\sigma(k)}|a_{\sigma(k+n)}).
\label{generalt}
\end{eqnarray}
The sum is over the $(2n-1)!!$ permutations $\sigma$
of $\{1,\dots,2n\}$ such that
$\sigma(1) < \sigma(2) < \cdots < \sigma(n)$
and $\sigma(k) < \sigma(n+k)$ for all $k=1,\dots,n$.
The proof is given by Caianiello \cite{Caianiello}.
Notice that, if the $t$-map was a state, 
equation (\ref{generalt}) would show that it is a
quasifree state \cite{Kay1}.
 
The deeper reason behind the $T$-map is Hopf algebra cohomology. If
the Laplace pairing is symmetric it is an exact 2-cocycle and thus
arises as the boundary of a 1-cochain. This is $t$. Furthermore the
twist deformed
algebra (with twisted product) is then isomorphic to the original one
(with symmetric product) giving rise to the
algebra automorphism $T$. The detailed
explanation of these facts is deferred to \cite{BrouderQG}.

\section{Relation to quantum field theory}
\label{sec:qft}
 
In this section, we make a connection
between the Hopf algebra approach and the usual
quantum field formalism.
We first consider the algebra $V$ of field operators
$\phi(f)$, equipped with the normal product.
Then, we define a Laplace pairing such that the
twisted product reproduces the standard operator
product of quantum field theory.
In other words, we define the operator product 
as a deformation of the normal product.
Textbooks sometimes proceed in the opposite direction and
describe a kind of ``normal ordering operation'', denoted by $:\,\,
:$, which takes a product of operators and
puts all creation operators on the left and all annihilation
operators on the right (see e.g.\ \cite{Itzykson} p.111). 
However, such an operation would
not be well-defined (\cite{Ticciati} p.28).
For instance, $a^\dagger(\kbf)\cdot a(\qbf) +\delta(\kbf-\qbf)$
and $a(\qbf)\cdot a^\dagger(\kbf)$
are equal as operators but their normal products
obtained after the action of $:\,\, :$
are
$a^\dagger(\kbf)\cdot a(\qbf) +\delta(\kbf-\qbf)$
and $a^\dagger(\kbf)\cdot a(\qbf)$, which are different.
Thus, it is not consistent to consider a normal product
as obtained from the transformation of an operator product.

\subsection{Fock space and definition of $V$}
\label{sec:Fock}
 
We first need to define a vector space $V$ of operators,
from which the space of normal products $S(V)$ will
be constructed.
The elements of $V$ are defined as operators acting
on a Fock space. For scalar bosons, we take $M=\mathbb{R}^N$ 
($N=3$ for the usual space-time)
and we consider the complex Hilbert space
of square integrable functions on $M$: $H=L^2(M)$.
To define the symmetric Fock space, we follow
Ref.\cite{ReedSimonI}, p.53.
If $H^{\otimes n}=H\otimes\cdots\otimes H$ and 
$S_n$ acts on $H^{\otimes n}$ by
\begin{eqnarray*}
S_n\big(f_1(\kbf_1)\otimes\cdots\otimes f_n(\kbf_n)\big)
&=& \frac{1}{n!}
\sum_{\sigma\in\Scal_n} f_{\sigma(1)}(\kbf_1)\otimes\cdots\otimes
f_{\sigma(n)}(\kbf_n),
\end{eqnarray*}
where $\Scal_n$ is the group of permutations of $n$ elements, then
the symmetric Fock space is
\begin{eqnarray*}
\Fcal_s(H) &=& \bigoplus_{n=0}^\infty S_n H^{\otimes n},
\end{eqnarray*}
where $S_0 H^{\otimes 0}=\mathbb{C}$ and $S_1 H^{\otimes 1}=H$.
An element $|\psi\rangle$ of $\Fcal_s(H)$ can be written
\begin{eqnarray*}
|\psi\rangle &=&
 (\psi_0,\psi_1(\kbf_1),\dots,\psi_n(\kbf_1,\dots,\kbf_n),\dots),
\end{eqnarray*}
where $\psi_0\in\mathbb{C}$ and each $\psi_n(\kbf_1,\dots,\kbf_n)\in
L^2(M^n)$ is
left invariant under any permutation of the variables.
The components of $|\psi\rangle$ satisfy
\begin{eqnarray*}
||\,|\psi\rangle ||^2 &=& \langle \psi |\psi\rangle 
= |\psi_0|^2+\sum_{n=1}^\infty \int|\psi_n(\kbf_1,\dots,\kbf_n)|^2 \dd
\kbf_1\dots\dd \kbf_n < \infty.
\end{eqnarray*}
An element of $\Fcal_s(H)$ for which $\psi_n=0$ for all but finitely
many $n$ is called a finite particle vector. The set of finite particle
vectors is denoted by $F_0$.
 
We define now the annihilation operator $a(\kbf)$ by its action
in $\Fcal_s(H)$.
The action of the operator $a(\kbf)$ on a vector $|\psi\rangle$
is given by the coordinates $(a(\kbf)|\psi\rangle)_n$
of $a(\kbf)|\psi\rangle$ (\cite{ReedSimonII} p.218):
\begin{eqnarray}
(a(\kbf)|\psi\rangle)_n(\kbf_1,\dots,\kbf_n) &=&
\sqrt{n+1}\, \psi_{n+1}(\kbf,\kbf_1,\dots,\kbf_n).
\label{defa(f)}
\end{eqnarray}
The domain of $a(\kbf)$ is
$D_{\Scal}=\{|\psi\rangle\in F_0, 
\psi_n \in \Scal(M^n) \,\,{\mathrm{for}}\,\, n>0\}$,
where $\Scal(M^n)$ is the Fr\'echet space of functions of rapid 
decrease\footnote{We recall that a function of rapid decrease is an
infinitely differentiable complex-valued function such that
\begin{eqnarray*}
||f||_{\alpha,\beta} &=&
\sup_{x\in M} |x^\alpha D^\beta f(x)| < \infty,
\end{eqnarray*}
for all $\alpha,\beta \in I_+^N$, where
$I_+^N$ is the set of all $N$-tuples of nonnegative integers
(see Ref. \cite{ReedSimonI}, p.133).}
on $M$.
 
The adjoint of $a(\kbf)$ is not a densely defined operator
but $a^\dagger(\kbf)$ is well-defined as a quadratic form
on $D_{\Scal}\times D_{\Scal}$: if $|\varphi\rangle$ and $|\psi\rangle$ belong
to $D_{\Scal}$ then (\cite{ReedSimonII}, p.219)
\begin{eqnarray*}
\langle \varphi| a^\dagger(\kbf) \psi\rangle &=&
\langle a(\kbf)\varphi| \psi\rangle,
\end{eqnarray*}
 
If $|\psi\rangle\in D_{\Scal}$, it can be checked that
$a(\kbf)|\psi\rangle\in D_{\Scal}$. Thus, we can calculate
the operator product $a(\kbf_1)\cdot a(\kbf_2)|\psi\rangle\in D_{\Scal}$
(the operator product is denoted by a dot $\cdot$).
The definition (\ref{defa(f)}) of $a(\kbf)$ and the
fact that the components of $|\psi\rangle$ are
symmetric enable us to prove that 
$a(\kbf_1)\cdot a(\kbf_2)=a(\kbf_2)\cdot a(\kbf_1)$.
Therefore $a^\dagger(\kbf_1)\cdot a^\dagger(\kbf_2)=
a^\dagger(\kbf_2)\cdot a^\dagger(\kbf_1)$
but nothing is known about
$a(\kbf_1)\cdot a^\dagger(\kbf_2)$, because $a^\dagger(\kbf_2)$ is defined
as a quadratic form, which means that all the
$a^\dagger(\kbf)$ must be on the left of all the $a(\kbf)$.
 
The vectors
$a^\dagger(\kbf_1)\cdot\dots\cdot a^\dagger(\kbf_m)
\cdot a(\qbf_1)\cdot\dots\cdot a(\qbf_n)$
for $m\ge0$ and $n\ge0$ generate a vector space denoted by $W$.
The element given by  $n=m=0$ is 1, the unit operator.
The space $W$ is equipped with a product, called
the normal product and denoted by $\pvee$.
If $u= a^\dagger(\kbf_1)\cdot\dots\cdot a^\dagger(\kbf_m)
\cdot a(\qbf_1)\cdot\dots\cdot a(\qbf_n)$
and $v\in W$, the normal product of $u$ and $v$ is
\begin{eqnarray*}
u\pvee v &=& v\pvee u = 
a^\dagger(\kbf_1)\cdot\dots\cdot a^\dagger(\kbf_m)
\cdot v  \cdot a(\qbf_1)\cdot\dots\cdot a(\qbf_n).
\end{eqnarray*}
This product is well defined because the creation operators
commute as well as the annihilation operators.
$W$ is a commutative and associative $*$-algebra with unit,
generated by $a(\kbf)$ and $a^\dagger(\kbf)$.
The $*$-structure is defined by
$a(\kbf)^*=a^\dagger(\kbf)$.
 
Now, we are ready to define the vector space $V$.
A smoothed field operator is an operator $\phi(f)$
defined by
\begin{eqnarray*}
\phi(f) &=& \int \dd x \int \frac{\dd\kbf}{(2\pi)^3\sqrt{2\omega_k}}
\Big(
f(x) \ee^{-i p\cdot x} a(\kbf) + f(x) \ee^{i p\cdot x} a^\dagger(\kbf)
\Big),
\end{eqnarray*}
where $\omega_k=\sqrt{m^2+|\kbf|^2}$ and $p=(\omega_k,\kbf)$.
The vector space $V$ is
\begin{eqnarray*}
V &=& \{\phi(f), f\in \Dcal(\mathbb{R}^{N+1})\},
\end{eqnarray*}
where $\Dcal(\mathbb{R}^{N+1})$ is the space of infinitely differentiable
functions with compact support in $\mathbb{R}^{N+1}$.
$V$ is indeed a vector space, with
$\phi(f)+\phi(g)=\phi(f+g)$ and $\lambda \phi(f)=\phi(\lambda f)$.
These smoothed fields can also be written
$\phi(f)=\int \dd x f(x) \phi(x)$, where
\begin{eqnarray*}
\phi(x) &=& \int \frac{\dd\kbf}{(2\pi)^3\sqrt{2\omega_k}}
\Big(\ee^{-i p\cdot x} a(\kbf) + \ee^{i p\cdot x} a^\dagger(\kbf)\Big).
\end{eqnarray*}
The field operators $\phi(x)$ are defined as quadratic forms
(see \cite{ReedSimonII}, p. 223). They
look like a basis of $V$ but $\phi(x)$ does not belong to $V$.

\subsection{The Hopf algebra $S(V)$ of normal products}
\label{normalprodsect}
 
Now that we have the vector space $V$, the Hopf algebra $S(V)$ is
defined according to Section~\ref{sec:symHopf}.
However, we want to identify elements of $S(V)$ (and not only $V$) as
operators in the sense of Section~\ref{sec:Fock}.
We know that $S^0(V)=\mathbb{C} 1$, where $1$ is the unit operator:
for any $|\psi\rangle \in F_s(H)$, $1|\psi\rangle=|\psi\rangle$.
We also know that $S^1(V)=V$. To identify $S^n(V)$ as a space of
operators in the desired sense we only
need to deduce the symmetric product of $S^n(V)$ from the
normal product of creation and annihilation operators. 
For example
\begin{eqnarray*}
\phi(f)\pvee \phi(g) &=& \int \dd x \dd x'\int
\frac{\dd\kbf}{(2\pi)^3\sqrt{2\omega_k}}
\frac{\dd\kbf'}{(2\pi)^3\sqrt{2\omega_{k'}}}
f(x) g(x') 
\\&&\hspace*{-1mm}\times
\Big(\ee^{-i (p\cdot x+ p'\cdot x')} a(\kbf)\pvee a(\kbf') 
+ \ee^{-i (p\cdot x- p'\cdot x')} a(\kbf) \pvee a^\dagger(\kbf')
\\&&\hspace*{-0mm}
+ \ee^{i (p\cdot x- p'\cdot x')} a^\dagger(\kbf)\pvee a(\kbf')
+ \ee^{i (p\cdot x+ p'\cdot x')} a^\dagger(\kbf)\pvee a^\dagger(\kbf')
\Big)\\
&=& \int \dd x \dd x'\int
\frac{\dd\kbf}{(2\pi)^3\sqrt{2\omega_k}}
\frac{\dd\kbf'}{(2\pi)^3\sqrt{2\omega_{k'}}}
f(x) g(x') 
\\&&\hspace*{-1mm}\times
\Big(\ee^{-i (p\cdot x+ p'\cdot x')} a(\kbf)\cdot a(\kbf') 
+ \ee^{-i (p\cdot x- p'\cdot x')} a^\dagger(\kbf') \cdot a(\kbf)
\\&&\hspace*{-0mm}
+ \ee^{i (p\cdot x- p'\cdot x')} a^\dagger(\kbf)\cdot a(\kbf')
+ \ee^{i (p\cdot x+ p'\cdot x')} a^\dagger(\kbf)\cdot a^\dagger(\kbf')
\Big).
\end{eqnarray*}
In the above expression, the integral over $x$ and
$x'$ gives the Fourier transform of $f$ and $g$, which is
square integrable because of Plancherel's theorem (\cite{ReedSimonII} p.4).
Thus, theorem X.44 of Ref.\ \cite{ReedSimonII} p.220 can be applied
to show that the product is a well defined operator on $D_{\Scal}$.
The same reasoning shows that the symmetric product $\pvee$ is
well defined in $S(V)$.
By definition the coproduct of $S(V)$ is generated by
\begin{eqnarray*}
\Delta \phi(f) &=& \phi(f)\otimes 1+ 1\otimes \phi(f).
\end{eqnarray*}
The antipode is defined 
by $\antip(\phi(f_1)\pvee\dots\pvee\phi(f_n)=
(-1)^n \phi(f_1)\pvee\dots\pvee\phi(f_n)$.
 
The counit 
has a direct quantum field meaning: it is the expectation value
over the vacuum.
We first define a vector of $D_\Scal$ called the vacuum:
$|0\rangle=(1,0,0,\dots)$. From the definition (\ref{defa(f)}) of
$a(\kbf)$ we see that $a(\kbf)|0\rangle=0$ and $\langle 0|a^\dagger(\kbf)=0$.
Therefore, $\langle 0| u |0\rangle = 0 =\varepsilon(u)$,
for any $u\in S^n(V)$ if $n>0$.
Moreover, 
$\langle 0| 1 |0\rangle = \langle 0| 0\rangle=1=\varepsilon(1)$.
Thus, the map that sends
$u\in S(V)$ to $\langle 0| u |0\rangle$ is a linear
map from $S(V)$ to $\mathbb{C}$ which is equal to 
$\varepsilon$ on $S^n(V)$ for all $n$.
Therefore, $\varepsilon(u)=\langle 0| u |0\rangle$.
This identity is an additional argument in favor of the fact that
Hopf algebras are a natural framework for quantum field theory.
 
The Hopf algebra $S(V)$ has a $*$-structure (see appendix)
defined by
$(\lambda 1)^*=\bar\lambda 1$ for $\lambda\in\mathbb{C}$,
$\phi(g)^*=\phi(\bar g)$ for $a(g)\in V$,
and $(u\pvee v)^*=v^* \pvee u^*=u^* \pvee v^*$. This makes $S(V)$ a
Hopf $*$-algebra because $*$ is an antilinear involution
and a homomorphism of coalgebra:
\begin{eqnarray*}
\Delta (\phi(g)^*) &=& \Delta \phi(\bar g) =
\phi(\bar g)\otimes 1+ 1\otimes \phi(\bar g)
\\&=& \phi(g)^*\otimes 1+ 1\otimes \phi(g)^*=(\Delta \phi(g))^*,
\end{eqnarray*}
so that $\Delta u^* = \sum {u\i1}^*\otimes {u\i2}^*$
for any $u\in S(V)$.
Moreover, the $*$-structure respects the grading, 
because $u^*\in S^n(V)$ if $u\in S^n(V)$ and
the condition on the antipode is satisfied (\cite{Kassel} p.87):
for $u\in S^n(V)$ we have $S(u)=(-1)^n u$ and $S(u^*)=(-1)^n u^*$, thus
\begin{eqnarray*}
{S\big(S(u)^*\big)}^* &=& (-1)^n {S(u^*)}^*={(u^*)}^*=u.
\end{eqnarray*}
Thus, $S(V)$ is a Hopf $*$-algebra.
 
\subsection{Alternative constructions}
Our definition of $S(V)$ as operators acting on a Fock space 
is the most standard one but other methods exist to define $S(V)$.
For instance, we can consider an abstract Borchers-Uhlmann 
algebra \cite{BFK,Sahlmann,Strohmaier} equipped with a state. 
The GNS construction then yields a representation of
the algebra as operators acting on a Hilbert space.
The Borchers-Uhlmann approach is 
interesting because it can be used in a curved space-time.
 
\subsection{Green functions and propagators}
 
The Laplace pairings corresponding to operator product
and time-ordered product are to be defined from
various Green functions and propagators.
As many conflicting conventions can be found
for these in the literature,
we give now explicit expressions
for the functions we use in the case of the free
scalar field. We consider a flat four-dimensional space-time
with diagonal metric $(1,-1,-1,-1)$.
The following formulas are compiled from
Refs.\cite{Itzykson}, p. 33 and p.133,
\cite{Pauli}, Sections 5 and 13,
\cite{Scharf}, Section 2.3 and \cite{SchwingerI} Section 2-1,
after conversion to our notation.
 
The positive frequency function is
\begin{eqnarray*}
G_+(x) &=& \int \frac{d\kbf}{(2\pi)^3 2\omega_k} \ee^{-ip\cdot x}
= \int \frac{d^4p}{(2\pi)^3} \delta(p^2-m^2) \theta(p^0)\ee^{-ip\cdot x},
\end{eqnarray*}
where $\omega_k=\sqrt{m^2+|\kbf|^2}$ and $p=(\omega_k,\kbf)$.
The symmetric function is
\begin{eqnarray*}
G^1(x) &=& (1/2)\big(G_+(x)+G_+(-x)\big)
= \int \frac{d\kbf}{(2\pi)^3 2\omega_k} \cos(p\cdot x)
\\&=& \int \frac{d^4p}{2(2\pi)^3} \delta(p^2-m^2) \ee^{-ip\cdot x}.
\end{eqnarray*}
The commutator is 
\begin{eqnarray*}
G(x) &=& -i\big(G_+(x)-G_+(-x)\big)
= -\int \frac{d\kbf}{(2\pi)^3\omega_k} \sin(p\cdot x)
\\&=& -i\int \frac{d^4p}{(2\pi)^3} \delta(p^2-m^2) \signe(p^0)\ee^{-ip\cdot x}.
\end{eqnarray*}
The commutator is related to the retarded Green function by
$G_\mathrm{ret}(x)=-\theta(x^0) G(x)$ and to the advanced one by
$G_\mathrm{adv}(x)=\theta(-x^0) G(x)$.
Finally, the Stueckelberg-Feynman Green function is
\begin{eqnarray*}
G_F(x) &=& iG_+(x) + G_\mathrm{adv}(x)=
 -\int \frac{d^4p}{(2\pi)^4} \frac{\ee^{-ip\cdot x}}{p^2-m^2+i\epsilon}.
\end{eqnarray*}
The functions $G_\mathrm{adv}(x-y)$, $G_\mathrm{ret}(x-y)$
and $G_F(x-y)$ satisfy 
$(\partial_\mu\partial^\mu+m^2) G_X=\delta(x-y)$, the functions
$G_+(x-y)$, $G(x-y)$ and $G^1(x-y)$ 
satisfy $(\partial_\mu\partial^\mu+m^2) G_X=0$.
In the case of a massless scalar field, these functions are
\begin{eqnarray*}
G_+(x) &=& -\frac{1}{4\pi^2}\big( P\frac{1}{x^2} +i\pi\,
                \signe(x^0)\delta(x^2)\big),\\
G^1(x) &=& -\frac{1}{4\pi^2} P\frac{1}{x^2},\quad
G(x) = -\frac{1}{2\pi}\signe(x^0)\delta(x^2),\\
G_\mathrm{ret}(x) &=& \frac{1}{2\pi}\theta(x^0)\delta(x^2),\quad
G_\mathrm{adv}(x) = \frac{1}{2\pi}\theta(-x^0)\delta(x^2),\\
G_F(x) &=& -\frac{i}{4\pi^2} \frac{1}{x^2-i\epsilon}.
\end{eqnarray*}
 
The corresponding Green functions and propagators for
a real scalar field in curved space-time or in an
external potential can be found in Ref.\cite{Fuller}.
Notice, that $G_+(x)$ satisfies the Hadamard condition
(see, e.g.\ equation (3) in \cite{Radzikowski}) because 
it can be written as
\begin{eqnarray*}
G_+(x) &=& \frac{1}{(2\pi)^2}\,\frac{1}{-x^2+2i\varepsilon x^0 + \epsilon^2}.
\end{eqnarray*}

\subsection{Operator product}
In this section, we show that the choice of a 
Laplace pairing enables us to define a twisted product
that reproduces the usual operator product of quantum field theory.
 
\subsubsection{Laplace pairing}
\label{LPsect}
The twisted product is defined from the Laplace pairing
\begin{eqnarray}
(\varphi(f)|\varphi(g))_+ &=& \int \dd x \dd y f(x) G_+(x-y) g(y).
\label{Gplus}
\end{eqnarray}
by (\ref{deftwisted}). We denote it by $\oprod$. For example
\begin{eqnarray*}
\varphi(f)\oprod\varphi(g) &=& \varphi(f)\pvee\varphi(g) +
(\varphi(f)|\varphi(g))_+ .
\end{eqnarray*}
On the other hand, Wick's theorem gives the operator
product as the sum of the normal product plus the average
value over the vacuum:
\begin{eqnarray*}
\varphi(f)\cdot\varphi(g) &=& \varphi(f)\pvee\varphi(g) +
\langle 0 | \varphi(f)\cdot \varphi(g) | 0 \rangle.
\end{eqnarray*}
The last term can be written
$\langle 0 | \varphi(f)\cdot \varphi(g) | 0 \rangle =
\int \dd x \dd y f(x) g(y)
\langle 0 | \varphi(x)\cdot \varphi(y) | 0 \rangle$, where
$\langle 0 | \varphi(x)\cdot \varphi(y) | 0 \rangle$
is the Wightman function $G_+(x-y)$. Thus, equation (\ref{Gplus})
shows that $\varphi(f)\oprod\varphi(g)=\varphi(f)\cdot\varphi(g)$
and the twisted product of two fields is equal to
their operator product.
 
\subsubsection{Wick's theorem}
We have shown that the operator product of two field operators can
be defined as a twisted product. We must show now that this
is still true for the product of any number of field operators.
As the relation between operator product and normal product is
governed by Wick's theorem, it is enough to show
that Wick's theorem is satisfied 
in $S(V)$. For a related approach in the fermionic
case, see Reference~\cite{Fauser}.
 
Wick's theorem is very well known, so we recall it only briefly.
It states that the operator
product of a given number of elements of $V$
is equal to the sum over all possible pairs of
contractions (see e.g. \cite{Fetter} p.209,
\cite{WeinbergQFT} p.261, \cite{Ticciati} p.85).
A contraction $a^\bullet b^\bullet$
is the difference between the operator
product and the normal product. In Wick's
notation \cite{Wick}
$a^\bullet b^\bullet= ab-{:}ab{:}$.
This pairing was used by Houriet and Kind
even before Wick's article \cite{Houriet}.
The contraction is a scalar.
To express it in our notation, we identify the
normal product ${:}ab{:}$ with the symmetric
product $a\pvee b$. 
 
Wick's theorem is proved recursively from the following identity
\cite{Wick}:
\begin{eqnarray}
{:}a_1\dots a_n{:}\,b &=& {:}a_1\dots a_n b{:}+
\sum_{j=1}^n a_j^\bullet b^\bullet\, {:}a_1\dots a_{j-1}a_{j+1}\dots a_n{:}
\label{Wickid}
\end{eqnarray}
Thus, we must show that the above identity is valid for
the twisted product. In our notation, (defining $u=a_1\pvee\dots\pvee a_n$)
we must prove
\begin{eqnarray}
u\oprod b &=& u \pvee b
+ \sum_{j=1}^n (a_j|b)\,a_1\pvee\dots\pvee a_{j-1}\pvee
a_{j+1}\pvee\dots\pvee a_n.
\label{ucircb}
\end{eqnarray}
To show this, we use the definition
(\ref{deftwisted}) of the twisted product and equation (\ref{Deltaa}) to find
\begin{eqnarray}
u\oprod b &=& u \pvee b + \sum (u\i1|b) u\i2.
\label{ucircb2}
\end{eqnarray}
The Laplace pairing $(u\i1|b)$ is zero if the degree
 of $u\i1$ is different from 1.
In other words, $u\i1$ must be an element of $V$.
According to the general equation (\ref{shuffle}) for $\Delta u$, this
happens only for the $(1,n-1)$-shuffles.
By definition, a $(1,n-1)$-shuffle is a permutation
$\sigma$ of $(1,\dots,n)$ such that
$\sigma(2)<\dots <\sigma(n)$, and the corresponding
terms in the coproduct of $\Delta u$
are
\begin{eqnarray*}
\sum_{j=1}^n a_j \otimes a_1\pvee\dots\pvee a_{j-1}\pvee
a_{j+1}\pvee\dots\pvee a_n.
\end{eqnarray*}
Thus, we recover (\ref{ucircb}) from the Laplace identity
(\ref{ucircb2}).
Equation (\ref{ucircb}) is equivalent to Wick's identity (\ref{Wickid})
because
the symmetric product of equation (\ref{ucircb}) is the normal product
of equation (\ref{Wickid}), and the Laplace pairing of equation (\ref{ucircb}) 
is the same as the contraction of equation (\ref{Wickid}).
Hence, the left-hand sides are identical and the twisted product 
of $u$ and $b$ is the operator product of $u$ and $b$.
From the associativity of the operator and twisted products, we deduce that
the twisted product of any number of operators is equal to their
operator product. Therefore, the twisted product is equal to the operator
product on $S(V)$.
 
Wick's theorem enables us to calculate the operator product of elements of $V$.
The quantum group approach enables us to express the operator product of
elements of $S(V)$ as
\begin{eqnarray}
u_1\oprod \dots \oprod u_n &=&
\Big(
\prod_{i=1}^{n-1} \prod_{j=i+1}^{n} ({u_i}_{(j-1)}|{u_{j}}_{(i)})
\Big)
u_{1(n)}\pvee\dots\pvee u_{n(n)},
\label{generWick}
\end{eqnarray}
where $u_{i(j)}$ is obtained by composing $n-1$ times the coproduct of $u_i$.
For instance, for $n=4$,
\begin{eqnarray*}
(\Id\otimes\Id \otimes \Delta)(\Id \otimes \Delta) \Delta u
&=& \sum u\i1 \otimes u\i2 \otimes u\i3 \otimes u\i4.
\end{eqnarray*}
A simple example of this generalized Wick's theorem is
\begin{eqnarray*}
u_1 \oprod u_2 \oprod u_3
&=&
\sum ({u_1}\i1|{u_{2}}\i1)({u_1}\i2|{u_{3}}\i1)({u_2}\i2|{u_{3}}\i2)
 {u_{1}}\i3 \pvee {u_{2}}\i3 \pvee {u_{3}}\i3.
\end{eqnarray*}
A proof of equation (\ref{generWick}) for the general case
of $S(C)$, where $C$ is a cocommutative coalgebra is given
in \cite{BrouderGroup24}.

\subsection{Time-ordered product}
 
In this section, we show that the choice of a 
Laplace pairing enables us to define a twisted product
that reproduces the usual time-ordered product of quantum field theory.
 
We start with the Laplace pairing
\begin{eqnarray}
(\varphi(f)|\varphi(g))_F &=& -i\int \dd x \dd y f(x) G_F(x-y) g(y).
\label{LPTOP}
\end{eqnarray}
We denote by $\circ$ the corresponding twisted product.
For example
\begin{eqnarray*}
\varphi(f)\circ\varphi(g) &=& \varphi(f)\pvee\varphi(g) +
(\varphi(f)|\varphi(g))_F .
\end{eqnarray*}
In particular, it can be checked that
\begin{eqnarray*}
\langle 0 | \varphi(f)\circ\varphi(g) | 0 \rangle &=&
(\varphi(f)|\varphi(g))_F = -i\int \dd x \dd y f(x)G_F(x-y)g(y).
\end{eqnarray*}
As in section \ref{LPsect}, we see that the twisted product of two
operators is equal to their time-ordered product because
$-iG_F(x-y) = \langle 0|T\big(\varphi(x)\varphi(y)\big)|0\rangle$
(\cite{ReedSimonII} p.214).
The fact that the time-ordered product of any number of
field operators is equal to their twisted product with
Laplace pairing (\ref{LPTOP}) follows again from Wick's theorem
and is proved as for the operator product.
Notice that $G_F(x)$ is a symmetric distribution
(i.e. $G_F(-x)=G_F(x)$), thus the Laplace coupling
$(|)_F$ is symmetric and the time-ordered product is
commutative.
It might be useful to stress here that the twisted product
is always associative. This implies that the time-ordered
product is an associative and commutative product.
 
The role of the counit as the vacuum expectation value thus enables us
to express (smeared) correlation functions of free scalar quantum field
theory. For example, the $n$-point function smeared with test functions
$f_1, \dots, f_n$ is given by
\begin{eqnarray*}
 \langle 0 | T(\varphi(f_1) \varphi(f_2) \cdots \varphi(f_n)) | 0
 \rangle\\
 = \cou \left(\varphi(f_1)\circ\varphi(f_2)\circ
 \cdots\circ\varphi(f_n)\right)\\
 = t\left(\varphi(f_1)\pvee\varphi(f_2)\pvee
 \cdots\pvee\varphi(f_n)\right) ,
\end{eqnarray*}
where $t$ is defined from $\circ$ as in Section~\ref{sec:tmap}.
The time-ordered product is particularly important
when we consider interacting quantum fields
because it enters the definition of the S-matrix
and the interacting Green functions.
In a forthcoming publication, it will be shown that
quantum groups methods allow for an easy handling of
the S-matrix and provide short algebraic proofs of the 
linked-cluster and coupled-cluster theorems.
 
\section{Conclusion}
 
The quantum group approach to quantum field theory
provides a second quantization without commutators.
To this end the algebra of field operators with normal product is
equipped with a Hopf algebra structure. Green functions are encoded
in the Laplace pairing.
Both operator product and time-ordered product arise as twisted
products from Laplace pairings via Hopf algebra cohomology theory.
The relationship
between normal, operator and time-ordered product is thus completely
encoded in the algebraic expressions. In particular, this is also true
for Wick's theorem. Combinatorial structures and proofs in quantum
field theory are considerably simplified by casting them in closed
algebraic expressions.
 
On the other hand, the approach advocated here brings second
quantization much closer in spirit to deformation quantization. If we
think of the field operators with normal product as classical
observables (arising from classical field configurations as in the
path integral picture) their operator product as quantum observables
arises indeed as a twist deformation of the classical commutative
product \cite{Dito90,Dito92,Dutsch}.
 
Although we have considered here only free fields,
the formalism extends naturally to the interacting case
\cite{BrouderOeckl,BrouderGroup24}.
If we consider the exterior algebra $\Lambda(V)$ 
instead of the symmetric algebra $S(V)$ we can
carry out the second quantization of fermion fields
\cite{BrouderQG}.
 
While in this paper the Laplace pairing was considered to be scalar,
it is also possible to give it a value in an algebra.
This enables us to formally second-quantize associative
algebras or to consider matrix-valued states,
which should be useful in quantum field calculations
involving degenerate vacua.
Moreover, the quantum group approach provides powerful
algorithms to calculate quantum field theory expressions
with correlated vacua, which are used
in solid-state physics and quantum 
chemistry \cite{BrouderOeckl,Kutzelnigg}.

\ack
We are extremely grateful to Alessandra Frabetti
for her constant help and for her many suggestions
and clarifications. This paper would have been
much worse without her contribution.
We are also very grateful to Bertfried Fauser for
the detailed e-mails he sent us to
explain his view of the connection between
the twisted product and Wick's theorem, and
for the comments and corrections he made on the manuscript.
We thank Moulay Benameur and Georges Skandalis for very useful discussions
and Gudrun Pinter, Philippe Sainctavit, 
Eric-Olivier Lebigot for their precious comments.
R.~O.\ was supported by a Marie Curie fellowship of the European Union.
This is IPGP contribution \#0000.
 
\appendix
\section*{Appendix}
\setcounter{section}{1}
 
In this appendix, we make a short presentation of
Hopf algebras.
 
A {\sl {Hopf algebra}} is an algebra $\Acal$ equipped with
a coassociative coproduct $\Delta$, a counit $\varepsilon$ and an
antipode $\antip$,
such that $\Delta$ and $\varepsilon$ are algebra homomorphisms.
We now introduce these concepts step by step.
 
First, an {\sl {algebra}} is a vector space $\Acal$ over $\mathbb{C}$ 
equipped with an associative linear product over $\Acal$,
denoted $\cdot$ and a unit, denoted $1$.
A product is {\sl {linear}} if for any $a, b, c \in \Acal$:
$a\cdot(b+c)=a\cdot b+ a\cdot c$,
$(a+b)\cdot c= a\cdot c + b \cdot c$ and
$a\cdot (\lambda b) = \lambda (a\cdot b) = (\lambda a)\cdot b$.
It is {\emph{associative}} if for any $a, b, c \in \Acal$:
$(a\cdot b)\cdot c= a\cdot (b\cdot c)$.
The {\emph{unit}} is the element of $\Acal$ such that, for any $a\in \Acal$,
we have $1\cdot a=a\cdot 1=a$. A good example of an algebra
is the symmetric algebra $S(V)$ described in section \ref{sec:symHopf}.
 
The more unusual concept in a Hopf algebra is the
{\emph{coproduct}}, which is a linear map
from $\Acal$ to $\Acal\otimes\Acal$, denoted by $\Delta$.
A coproduct can often be considered as giving all the ways to split an
element of $\Acal$ into two ``parts''.
For the example of the symmetric algebra $S(V)$, the coproduct of
$x_1\pvee\dots\pvee x_n$ (with all $x_i$ in $V$) is defined as
\begin{eqnarray*}
\Delta x_1\pvee\dots\pvee x_n
&=& \sum_{I} P(I)\otimes P(I^c),
\end{eqnarray*}
where the sum is over all subsets $I$ of 
$\{x_1,\dots,x_n\}$, $I^c=\{x_1,\dots,x_n\} \backslash I$
and $P(I)$ is the symmetric product of the elements of $I$.
For instance,
\begin{eqnarray*}
\Delta 1 &=& 1\otimes 1, \\
\Delta x &=& x \otimes 1 + 1\otimes x,\\
\Delta x\pvee y &=& x \pvee y \otimes 1 +  x \otimes y
+ y \otimes x+ 1\otimes x\pvee y,\\
\Delta x\pvee y \pvee z &=& x \pvee y \pvee z \otimes 1 
+  x \otimes y \pvee z
+  y \otimes x \pvee z
+  z \otimes x \pvee y
\\&&
+ x \pvee y\otimes z
+ x \pvee z\otimes y
+ y \pvee z\otimes x
+ 1\otimes x \pvee y \pvee z.
\end{eqnarray*}
 
To recall this `splitting', the action of the coproduct is
denoted by $\Delta a = \sum a\i1 \otimes a\i2$,
where $a\i1$ and $a\i2$ are the first and second ``parts''
of $a$.
For instance, if $a= x  $, we have
$a\i1=x$ and $a\i2=1$ for the first
term of $\Delta a$ and 
$a\i1=1$ and $a\i2=x$ for its second term.
This is called Sweedler's notation.
 
In a Hopf algebra, the coproduct is compatible with
the product, in other words
$\Delta (a\cdot b) = (\Delta a)\cdot (\Delta b)$.
Or, more precisely
$\Delta (a\cdot b) = \sum (a\i1\cdot b\i1) \otimes (a\i2\cdot b\i2)$.
In mathematical terms, we say that the coproduct is an {\emph{algebra
homomorphism}}.
 
This compatibility is often used to define the coproduct. For instance, 
we can consider the algebra $\Jcal$ generated by the angular momentum operators
$J_x$, $J_y$ and $J_z$
(i.e. the universal enveloping algebra of $su_2$).
The product $\cdot$ is the operator product
and the coproduct is defined by the action of the angular momentum
operators on products of two states $|\psi_1\rangle\otimes|\psi_2\rangle$
\begin{eqnarray}
\Delta J_i &=& J_i \otimes 1 + 1 \otimes J_i. \label{DeltaJ}
\end{eqnarray} 
We recognize here the coupling of angular momenta.  
This gives us a physical meaning of the coproduct: it is a way
to go from a one-particle operator to a two-particle operator.
For products of operators $J_{i_1}\cdot \dots \cdot J_{i_n}$ the coproduct is
defined recursively by its compatibility with the operator product:
\begin{eqnarray*}
\Delta (J_{i_1}\cdot J_{i_2}\cdot \dots \cdot J_{i_n})
&=& \Delta (J_{i_1}) \cdot \Delta (J_{i_2}\cdot \dots \cdot J_{i_n}).
\end{eqnarray*}
For example
$\Delta (J_x \cdot J_y) = J_x \cdot J_y \otimes 1 + J_x\otimes J_y 
+ J_y \otimes J_x + 1 \otimes J_x \cdot J_y.$
Now it is natural to go from a two-particle operator to
a three-particle operator by repeating the action of the
coproduct. However, since $\Delta a = \sum a\i1\otimes a\i2$
it is not clear on which side the second $\Delta$ should act:
on $a\i1$ or on $a\i2$? 
If $\Delta$ is applied on $a\i1$ we obtain
\begin{eqnarray*}
(\Delta\otimes \Id) \Delta a &=&
\sum (\Delta a\i1) \otimes a\i2 =
\sum a\i1\i1\otimes a\i1\i2 \otimes a\i2,
\end{eqnarray*}
where we have written 
$\Delta a\i1 = \sum a\i1\i1\otimes a\i1\i2$.
If $\Delta$ is applied on $a\i2$ we obtain
\begin{eqnarray*}
(\Id\otimes \Delta) \Delta a &=&
\sum a\i1\otimes (\Delta a\i2)
= \sum a\i1\otimes a\i2\i1 \otimes a\i2\i2.
\end{eqnarray*}
A very important property of 
Hopf algebras is that the result does not depend on which
side you apply the coproduct. In other word, for a Hopf
algebra we require that the coproduct satisfies 
$(\Delta\otimes \Id) \Delta a = 
(\Id\otimes \Delta) \Delta a$, i.e. 
\begin{eqnarray*}
\sum a\i1\i1\otimes a\i1\i2 \otimes a\i2
&=&
\sum a\i1\otimes a\i2\i1 \otimes a\i2\i2.
\end{eqnarray*}
This property is called the {\sl {coassociativity}} of the coproduct.
By applying again the coproduct, we can define the action of the angular
momentum operators on many-particle states. The coassociativity ensures
that the result does not depend on the order used.
If the coproduct is seen from the splitting point of view, 
coassociativity means that the splitting of an element into three 
parts does not depend on whether you further split the first or the 
second part.
It can be checked that the coproduct of the symmetric algebra 
$S(V)$ and of the universal enveloping algebra $\Jcal$ are coassociative.
 
We still need two ingredients to make a Hopf algebra: a {\emph{counit}} and
an {\emph{antipode}}.
A {\emph{counit}} is a linear map from $\Acal$ to $\mathbb{C}$, denoted
by $\varepsilon$, such
that $\sum a\i1\varepsilon(a\i2) = \sum\varepsilon(a\i1) a\i2=a$.
In a Hopf algebra, the counit $\varepsilon$ is an algebra homomorphism
(i.e. $\varepsilon(a\cdot b)=\varepsilon(a)\varepsilon(b)$).
In most cases, the counit is very simple. For $S(V)$ we have
$\varepsilon(1)=1$,
$\varepsilon(x_1\pvee \dots\pvee x_n)=0$ for $n>0$, 
for $\Jcal$ we have $\varepsilon(1)=1$,
$\varepsilon(J_{i_1}\cdot\dots\cdot J_{i_n})=0$ for $n>0$.
Finally the {\emph{antipode}} is a linear map from $\Acal$ to $\Acal$ such that
$\sum a\i1\cdot \antip(a\i2) = \sum \antip(a\i1)\cdot a\i2=\varepsilon(a) 1$.
The antipode is a kind of inverse. It is
anti-multiplicative, i.e.\  $\antip(a\cdot b)= \antip(b) \cdot \antip(a)$.
In our examples: 
for $S(V)$ the antipode is defined by
$\antip(1)=1$ and 
$\antip(x)=-x$, so that 
$\antip(x_1\pvee\dots\pvee x_n)=(-1)^n x_1\pvee \dots\pvee x_n$ for $n>0$,
for $\Jcal$ the antipode is defined by
$\antip(1)=1$ and
$\antip(J_{i})=-J_i$, so that 
$\antip(J_{i_1}\cdot\dots\cdot J_{i_n})=(-1)^n J_{i_n}\cdot\dots\cdot
J_{i_1}$ for $n>0$.
 
A Hopf algebra is {\emph{commutative}} if, for any $a,b \in \Acal$:
$a\cdot b= b\cdot a$. A Hopf algebra is
{\emph{cocommutative}} if, for any $a\in \Acal$:
$\sum a\i1\otimes a\i2= \sum a\i2 \otimes a\i1$,
in other words, if $a$ can be split into $a\i1$ and
$a\i2$, then there is also a splitting of $a$
where $a\i2$ is the first part and $a\i1$ is the second part.
The reader can check that $S(V)$ and $\Jcal$ are cocommutative
Hopf algebras. More generally, the universal enveloping
algebra of any Lie algebra is a cocommutative Hopf algebra.
 
The next concept we shall use is that of a graded Hopf algebra.
An algebra is {\sl {graded}} if there exist subspaces 
$({\Acal}_i)_{i\in\mathbb{N}}$ such that
$\Acal={\Acal}_0 \oplus {\Acal}_1\oplus{\Acal}_2\oplus\cdots$
and $ {\Acal}_m\cdot {\Acal}_n\subset {\Acal}_{m+n}$
for all $m,n\in\mathbb{N}$ (thus $1\in {\Acal}_0$).
An element $a$ of $\Acal$ is said to be \emph{homogeneous}
if there is an $m$ such that $a\in{\Acal}_m$. The 
\emph{degree} of $a$ is then $|a|=m$. 
A {\emph{graded Hopf algebra}} is a Hopf algebra which is 
graded as an algebra and such that
$\Delta {\Acal}_n \subset \bigoplus_{m=0}^n
{\Acal}_m\otimes{\Acal}_{n-m}$
and $\antip({\Acal}_n)\subset{\Acal}_n$ for all $n$.
If all elements of ${\Acal}_0$ 
can be written $\lambda 1$, the algebra is said to be {\sl {connected}}.
 
In our two examples, $S(V)$ is a connected graded Hopf algebra, where 
the degree is
given by $|x_1\pvee\dots\pvee x_n|=n$ but $\Jcal$ is not a 
(non-trivially) graded algebra
because of the commutator identity
$J_x\cdot J_y-J_y \cdot J_x=iJ_z$.
 
Finally, we define the concept of a Hopf $*$-algebra 
(\cite{Kassel} p.86 and \cite{Majid} p.31).
A $*$-algebra is an algebra $\Acal$ together with 
a map $*$ from $\Acal$ to $\Acal$ which is antilinear
($(a+b)^*=a^*+b^*$ and $(\lambda a)^*={\bar\lambda} a^*$,
where ${\bar\lambda}$ is the complex conjugate of $\lambda$)
and obeys ${(a^*)}^*=a$ and $(ab)^*=b^* a^*$
for any  $a$ and $b$ in $\Acal$.
A Hopf $*$-algebra $\Acal$ is a Hopf algebra which is
also a $*$-algebra and which satisfies the following compatibility
conditions:
$\Delta (a^*) = {(\Delta a)}^*$ (where we define
$(a\otimes b)^*= a^* \otimes b^*$),
$\varepsilon(a^*)=\overline{\varepsilon(a)}$
and
${S(S(a)^*)}^*=a$ for any $a\in\Acal$.
For more information on Hopf algebras and quantum groups, see 
Refs.\cite{Majid,Kassel,Chari,Dascalescu,Loday98}.
 
\section*{References}

\end{document}